\documentclass[12pt]{article} \usepackage{latexsym}
\usepackage{amsmath}

\begin{document}

\bibliographystyle{plain}

\def\one{{\bf 1}}
\def\dal{{\Box}}
\def\eps{{\epsilon}}
\def\cio{{\cal C}^\infty_0}
\def\ci{{\cal C}^\infty}
\def\Dp{{{\cal D}^{\prime}}}
\def\Sp{{{\cal S}^{\prime}}}
\def\Ep{{{\cal E}^{\prime}}}
\def\zero{{\bf 0}}
\def\k{{\bf k}}
\def\l{{\bf l}}
\def\wk{\omega_\k}
\def\lk{\lambda_\k}
\def\p{{\bf p}}
\def\bfsig{{\bf \sigma}}
\def\u{{\bf u}}
\def\x{{\bf x}}
\def\y{{\bf y}}
\def\z{{\bf z}}
\def\K{{\bf K}}
\def\R{{\bf R}}
\def\E{{\rm e}}
\def\ss{\mbox{ sing supp }}
\def\WF{\mbox{WF }}
\def\supp{\mbox{supp }}
\def\conesupp{\mbox{cone supp }}
\def\Srd{S_{\rho,\delta}}
\def\Lrd{L_{\rho,\delta}}
\def\Ird{I_{\rho,\delta}}
\def\half{{\frac{1}{2}}}
\def\GL{{\rm{GL}}}
\def\diag{{\rm diag}}
\def\real{{\rm Re}}

% TeX macros for Feynman slash notation
% (for linear combinations of Dirac matrices)

% Four slash macros:
% J\"org Knappen,   knappen@vkpmzd.kph.uni-mainz.de
% (\slashii is the TeXsis macro, \slashiii Rainer Sch"opf's one)

\def\slashi#1{\rlap{\sl/}#1}
% slashi works best for:
%a,c,e,f,g,h,k,o,r,s,u,v,x,y,z,I,J,S,T,V,Y,Z,\epsilon,\varepsilon,\partial
%
\def\slashii#1{\setbox0=\hbox{$#1$}             % set a box for #1
   \dimen0=\wd0                                 % and get its size
   \setbox1=\hbox{\sl/} \dimen1=\wd1            % get size of /
   \ifdim\dimen0>\dimen1                        % #1 is bigger
      \rlap{\hbox to \dimen0{\hfil\sl/\hfil}}   % so center / in box
      #1                                        % and print #1
   \else                                        % / is bigger
      \rlap{\hbox to \dimen1{\hfil$#1$\hfil}}   % so center #1
      \hbox{\sl/}                               % and print /
   \fi}                                         %
% \slashii works best for:
% b,i,l,n,t,w,x,y,z,A,B,E,L,Q,W
%
\def\slashiii#1{\setbox0=\hbox{$#1$}#1\hskip-\wd0\hbox
to\wd0{\hss\sl/\/\hss}}
% \slashiii works best for:
% d,f,j,l,\ell,m,w,x,z,C,D,F,G,H,K,L,M,N,O,P,Q,U,W,X,\Nabla,\partial
%
\def\slashiv#1{#1\llap{\sl/}}
% \slashiv works best for:
% e,g,p,q,y,z

\newcommand{\fig}[1]{Fig.~\ref{fig:#1}}
\newcommand{\figlabel}[1]{\label{fig:#1}}

  \title{Phase Space Factor for Two-Body Decay if One Product is a Stable Tachyon} 
  \author{Marek J. Radzikowski$^*$ ({\tt{radzik@physics.ubc.ca}}) \\
	\small{$^*$Present address: UBC Physics and Astronomy Dept.} \\ 
	\small{6224 Agriculture Road} \\
	\small{Vancouver, BC V6T 1Z1, Canada}}
  \date{\today}
  \maketitle
  %Dept. of Physics and Astronomy, U.B.C., Vancouver, B.C., V2T 2A6, Canada \\
   
\begin{abstract}
We calculate the phase space factor for a two-body decay in which one of the products 
is a tachyon. Two threshold conditions, a lower and an upper one, 
are derived in terms of the masses of 
the particles and the speed of a preferred frame. Implicit in the derivation is 
a consistently formulated quantum 
field theory of tachyons in which spontaneous Lorentz symmetry breaking occurs.
The result is to be contrasted with a parallel calculation by Hughes and 
Stephenson, which, however, implicitly adheres to strict Lorentz 
invariance of the underlying quantum field theory and produces the conclusion that there is
no threshold for this process.  
\end{abstract}

\baselineskip 2em

\section{Introduction}

We deal with a long-standing obstacle to the computation of the two-body
decay rate involving a tachyon as a product: the claim by \cite{HugSte90} that 
there is no threshold for the phase space factor $R_2$ when a particle of 
(regular) mass $M$ decays into a regular particle of mass $m_1$  and a 
tachyon of mass parameter $m_2$. Related to this is the observation that the formula 
obtained in \cite{HugSte90}
for $R_2$ (in which $2$ is a tachyon) is not obtained by the simple replacement
$m_2^2\longrightarrow -m_2^2$ into the usual formula for $R_2$ (in which $2$ 
is a regular particle). 

In this letter, we first point out that a scalar 
quantum field theory of tachyons cannot lead to a sensible description of 
particles if the usual Lorentz invariance axiom is insisted upon. Rather, we assume 
that an underlying cutoff exists in the one-particle energy-momentum spectrum
associated with the QFT. This property may be characterized
by stating that {\it a preferred (inertial) frame exists in which all the energies of both 
particle and anti-particle are positive or zero}. Lorentz 
transformation from the preferred frame then determines the $4$-momenta in the
energy-momentum spectrum in any other (inertial) frame; therefore, in any frame, boosted
with respect to the preferred one, the energy components of certain $4$-momenta are allowed 
to be negative. Clearly, this situation can be viewed as one in which 
spontaneous symmetry breaking of the Lorentz group has occurred. We may also refer 
the reader to \cite{CibRem96} for an alternative formulation
using an unconventional synchronization scheme, which also yields a preferred frame 
in the QFT.
  
We claim that with this cut-off in the one-particle spectrum, a lower threshold
depending on the speed of the preferred frame $\beta$ is introduced into the
two-body decay scenario. Furthermore, when this $\beta$ is fixed, and the 
tachyonic mass parameter $m_2$ is allowed to approach $0$, while $M,m_1$ are held fixed, 
the limit of $R_2$ obtained is the same as it would be if $m_2$ were a regular 
mass allowed to approach $0$. Furthermore, with $\beta$ fixed and $m_2$ small enough, the 
formula for $R_2({\rm tachyonic\ particle\ } 2)$ is indeed obtained from 
$R_2({\rm regular\ particle\ } 2)$ by the replacement $m_2^2 \longrightarrow -m_2^2$. 

By arguing that an underlying QFT involving Lorentz symmetry breaking should 
be used in calculating the phase space factors, and presumably could be used to make 
sense of any other QFT type calculation, we dispute the conclusion of 
\cite{HugSte90}, namely that ``there is strong circumstantial evidence against 
the proposal that at least one neutrino is a tachyon.'' In effect, therefore, 
the present letter provides indirect support for the 
{\it tachyonic neutrino hypothesis} of Chodos, Hauser, and Kosteleck\'y 
\cite{CHK85}, whose ideas the authors of \cite{HugSte90} were clearly
attempting to refute. 

A brief outline of the contents of this letter is as follows: In Section \ref{QFT} 
we summarize the basic elements which we deem essential in a tachyonic quantum 
field theory. We describe briefly why we consider the cutoff property to be 
necessary in the theory, and also offer hints as to how we expect it to be
sufficient for (a new notion of) causality and for renormalization. Section \ref{KinVar}
reviews the results of calculating the values of the energies of the product particles,
as well as the common magnitude of their momenta, first for the case of regular (massive)
product particles, then for the case that one is tachyonic. In Section \ref{ThreCond}, we
(re-)derive the single threshold condition for the regular mass case, and the lower 
and upper threshold conditions for the tachyonic case.
Section \ref{PhaFac} similarly covers the calculation for the 
two-body phase space factor for the two cases (regular and tachyonic). Section \ref{LimCase}
justifies the claims made in the introduction concerning the (good) behaviour of the phase space factor
in the tachyonic case when the tachyonic mass parameter goes to zero. In the conclusions
section (Section \ref{Conc}), we attempt to round out the list of clarifications and modifications
that would need to be applied to the parts of \cite{HugSte90} dealing with two-body decay involving 
a tachyon. In addition, we work out some of the details of this approach for pion decay,
and its reverse process, within the context of assuming that the muon neutrino is tachyonic.

\section{Rudiments of tachyonic quantum field theory}\label{QFT}

In the following we briefly explain the reasons why a quantum field
theory of tachyons is expected to require a preferred frame.

To start with, we choose units in which
$c=\hbar=1$ and the metric $\eta_{\mu\nu} = \diag(1, -1, -1, -1)$. Here,
$\Box = \eta^{\mu\nu}\partial_\mu \partial_\nu$. We assume that, when 
constructing a reasonable quantum field theory
of tachyons, a choice of space-like hyperplane (i.e., a 3-dimensional
subspace of 4-dimensional Minkowski space) through the origin in
energy-momentum space, labelled by $(E,\p)$, 
must be made to separate ``positive'' and ``negative'' energy 
modes. We note that, for the tachyonic case, these labels reduce to 
their usual meaning in the preferred
frame, but are mere labels for any inertial frame Lorentz boosted with 
respect to that frame. (Perhaps a better pair of terms for these would be ``upper'' and 
``lower''.) This distinction is easy to make in the
usual ($m^2\ge 0$) case, since the mode solutions of, say, the Klein-Gordon 
equation $\Box \phi + m^2 \phi = 0$ of the form
\begin{equation}\label{soln}
\E^{-i(Et-\p\cdot\x)} 
\end{equation}
have four-momenta $(E,\p)$ which lie on the
hyperboloid of two sheets $E^2 - \p^2 = m^2$, with $E>0$ or $E<0$. (Here
$E=0$ is included if $m=0$.) In the case of $m^2>0$, the upper sheet is
the natural choice to make to define positive energy
modes. One may regard any space-like hyperplane through the origin as
separating the upper and lower mass hyperboloids. 

In the case of $m=0$,
we delete the point $(E,\p)=0$ from the cone $E^2-\p^2=0$, and once again
any space-like hyperplane through the origin separates the upper and lower
cones.   

However, for the tachyonic case, the Klein-Gordon equation becomes 
$\Box\phi - m^2\phi = 0$, with a tachyonic mass parameter $m^2>0$, 
and the modes of the form (\ref{soln}) lie on the
hyperboloid of one sheet, $E^2 - \p^2 = -m^2$. Thus, the choice of which
modes are ``positive'' and which are ``negative'' (or ``upper'' and ``lower'')
energy depends upon the
choice of space-like hyperplane through the origin used to separate them. 
Hence, Lorentz symmetry is spontaneously broken at this point in the construction. 

The alternative to breaking Lorentz symmetry
would be to treat all of the modes (resp. none of them) 
as ``positive energy'' modes, and in any frame the
appearance of arbitrarily large negative particle energies 
(resp. large negative anti-particle energies) would lead to an
unsuitable theory from the QFT point of view. The problem is that the
singularities of the two-point function and Feynman propagator (the Green's
functions), which arise from a theory which is strictly Lorentz invariant (in the 
usual sense), would not be appropriate for constructing a
renormalizable theory when reasonable interactions are introduced. Specifically,
the {\it wave front set} of the theory would not satisfy a restriction (called elsewhere 
the ``wave front set spectral condition'' \cite{Rad92,Rad96a}, or the ``microlocal
spectrum condition'' \cite{BruFreKoeh96}) which, if satisfied, would ostensibly 
allow renormalization to proceed in a straightforward way \cite{BruFre00}.   

Insisting upon strict Lorentz invariance in the QFT would also introduce the 
possibility, however remote, of constructing causality-violating devices.
In other words, suitably well-equipped experimentalists would be enabled to 
construct a device consisting of a relay at a space-like separation from their own
lab, by which they could send a message backwards in time to themselves.
In progress (\cite{Rad06c}) is a more comprehensive discussion 
of these points, and a more detailed description of 
a model satisfying the above symmetry breaking requirement, which allows the 
wave front set condition on the two-point function to be satisfied,
renormalizability to be incorporated, and strong causality violations to be
circumvented. The model
would also exclude exponentially growing or decaying modes and so would avoid any 
possibility of producing unstable observables. Furthermore, it would exhibit a 
natural mechanism (i.e., at the basic QFT level) for maximal parity breaking of 
the neutrino. 

Now suppose a ``cut'' is made in the spectrum along $E'=0$
in a certain frame $\cal O'=\cal T$, so that positive and negative energy modes are
chosen to have $E'>0$ and $E'<0$ respectively. Frame $\cal T$ is denoted the {\it tachyon frame}.
In a lab frame $\cal O$, we find (according to the usual transformation law for a relativistic
boost) that a set of energies $E'$ labelled ``positive'' in the old 
frame $\cal T$ now has $E<0$ in
the lab frame $\cal O$. Furthermore, another set of modes labelled
``positive'' in $\cal T$ now no longer exists with those energies
in $\cal O$. From this one might argue that symmetry breaking still does 
not eliminate the ``unphysical'' negative energy 
modes from the theory. However, there exists an argument to rebut this: in $\cal O$ 
the negative energy particles are understood to be moving {\it backward} in time 
and can ultimately 
be ``re-interpreted'' as positive energy anti-particle states moving forward 
in time. Hence the usual notion of instability, in which particles which move {\it forward} 
in time with {\it negative} energy can be created out of the vacuum (such as may be 
encountered in situations in many body physics), is side-stepped here. 

For the purposes of clarity,
we shall not invoke the ``re-interpretation principle'' \cite{BilDesSud62} 
in $\cal O$, e.g., to ensure that the energies are always positive, but 
rather we shall consider the ``vacuum state'' as defining a particular designation of 
``upper'' (some of which are negative) and ``lower'' (some of which are 
positive) energies in $\cal O$. Perhaps there is some merit in viewing
the surface $E'=0$ in $\cal O'$ as defining the Fermi 
level of a half-filled Dirac sea of particles in $\cal O$, via a Lorentz 
transformation. (Note that this picture would be especially appropriate
for the spin-$\frac{1}{2}$ neutrino.) 

Also note that we shall assume the usual 
connection between spin and statistics, as do \cite{CKPG92}. Contrast 
Feinberg's assumption of fermionic statistics for the spin-$0$ case of 
tachyons \cite{Fei67}. We do not find that this assumption fits with the others 
we make here, since in the limit in which the tachyonic mass parameter goes to zero, 
the wrong connection of spin with 
statistics is obtained for a massless (scalar) theory.

\section{Dynamical variables for two-body decay}\label{KinVar}

The calculation for the decay rate of a particle of rest mass $M$ into two 
regular particles of rest masses $m_1$ and $m_2$, has been 
reviewed by \cite{HugSte90} and is a standard calculation in a first course in 
elementary particle physics \cite{Gri87}. We list the results here for 
convenience. The magnitude of the three momentum squared is 
\begin{equation}
   p^2\equiv p_1^2=p_2^2=\frac{(M^2-m_1^2-m_2^2)^2-4m_1^2m_2^2}{4M^2}\label{eqp1}
\end{equation} 
in the rest frame of $M$, while the energies of the resulting particles $1$ and $2$ are 
\begin{eqnarray}
   E_1 &=& \frac{M^2+m_1^2-m_2^2}{2M}\label{eq1}\;, \ \ \ {\rm and} \\
   E_2 &=& \frac{M^2-m_1^2+m_2^2}{2M}\label{eq2}\;,
\end{eqnarray} 
respectively. Eq.(\ref{eqp1}), simply by requiring $p^2$ to be positive, 
suggests the threshold inequality:
\begin{equation}
M\ge m_1+m_2 \;.   
\end{equation}

When $2$ is a tachyon, \cite{HugSte90} give a restriction they call 
the ``mass shell requirement'' which should read 
\begin{equation}
p^2\ge m_2^2\;, \label{MSR}
\end{equation} 
Here we have changed notation slightly from that of \cite{HugSte90}: in the 
tachyonic case, instead of allowing $m_2^2<0$ we replace $m_2^2$ by $-m_2^2$;
i.e., we change the sign in front of $m_2^2$ wherever this power of $m_2$
appears. E.g., $E_2^2-p_2^2=m_2^2<0$ becomes $E_2^2 - p_2^2 = -m_2^2<0$. Then 
$m_2$ is always understood to be a positive real parameter. Note that in 
Eq.(10) of \cite{HugSte90} the inequality goes the wrong way even with their
choice of sign convention for $m_2^2$. However the above 
formula Eq.(\ref{MSR}) follows easily from the formula for $p^2$ when $2$ is 
tachyonic:
\begin{eqnarray}
p^2 & = & \frac{(M^2-m_1^2+m_2^2)^2+4m_1^2m_2^2}{4M^2} \\
    & = & \frac{(M^2-m_1^2-m_2^2)^2+4M^2m_2^2}{4M^2} \ge m_2^2\label{eqp2}\;.
\end{eqnarray}
It appears the wrong inequality in Eq.(10) of \cite{HugSte90} is just a typo, 
so we have no quibble with it once it is corrected 
as in our formula Eq.(\ref{MSR}). In fact, it agrees with the notion
that there should be no exponentially growing modes (i.e., imaginary energy
modes) in the QFT. 

Note that when particle $2$ becomes tachyonic, Eqs.(\ref{eq1}) and (\ref{eq2}) 
become (by making the replacement $m_2^2\to-m_2^2$) 
\begin{eqnarray}
   E_1 &=& \frac{M^2+m_1^2+m_2^2}{2M}\label{eq3}\;, \ \ \ {\rm and} \\
   E_2 &=& \frac{M^2-m_1^2-m_2^2}{2M}\label{eq4}\;.
\end{eqnarray}
   	
\section{Threshold condition for tachyonic case}\label{ThreCond}

We now take into account the dependence of the lower bound of the tachyonic 
particle's energy $E_2$ on the preferred frame. We orient the axes so that the 
preferred frame is moving in the $-z$ direction, at speed $\beta$. The condition
that the lower bound of the particle's energy $E_2'$ is $0$ in the preferred frame
($E_2'\ge 0$) leads to the following condition in the rest frame of $M$:
\begin{equation}
   \gamma(E_2 + \beta p \cos\theta)\ge 0\;,
\end{equation}
where $\theta$ is the angle of particle $2$'s momentum $\p$ from the $z$ axis.  
Because of the mass shell condition $p=\sqrt{E_2^2 +m_2^2}$, this is equivalent to 
\begin{equation}
   E_2 \ge -\frac{m_2\beta\cos\theta}{\sqrt{1-\beta^2\cos^2\theta}}\label{eq5}\;.
\end{equation}
Combining Eqs.(\ref{eq4}) and (\ref{eq5}) we then obtain 
\begin{equation}
   \frac{M^2-m_1^2-m_2^2}{2M}\ge-\frac{m_2\beta\cos\theta}{\sqrt{1-\beta^2\cos^2\theta}}\;.
\end{equation}	
This inequality determines a range of values of $\theta$, which is possibly empty. These are the allowed 
directions of the momentum of the tachyon for which the decay can proceed. To find these,  
we express the formula $\frac{\cos\theta}{\sqrt{1-\beta^2\cos^2\theta}}$ as $\gamma g(x)$,
where $x=\cos\theta$, and note that 
\begin{equation}
   g(x)=\frac{x}{\gamma\sqrt{1-\beta^2x^2}}
\end{equation}
ranges between $-1$ and $1$ as $x$ varies from $-1$ to $1$. Thus, in order for the reaction 
to proceed for at least one direction, we must have 
\begin{equation}
   E_2 = \frac{M^2-m_1^2-m_2^2}{2M}\ge-m_2\beta\gamma\label{TC}\;.
\end{equation} 
This is the {\it lower threshold condition} for the case that particle $2$ is a tachyon, and the
preferred frame's speed is $\beta$ relative to the rest frame of $M$. This condition may also be
written as
\begin{eqnarray}
   (M+m_2\beta\gamma)^2 &\ge& m_1^2 + m_2^2\gamma^2\;, \ \ \ {\rm or} \\
   M &\ge& \sqrt{m_1^2 +m_2^2\gamma^2} - m_2\beta\gamma\;.
\end{eqnarray}
The range of directions for which decay is possible would then be
\begin{equation}
   1\ge x\ge -h\left(\frac{M^2-m_1^2-m_2^2}{2Mm_2\beta\gamma}\right)\;,
\end{equation}
where $h$ is the inverse function of $g$. Namely, 
\begin{equation}
   h(y)=\frac{\gamma y}{\sqrt{1+\beta^2\gamma^2 y^2}} = \frac{y}{\sqrt{1-\beta^2(1-y^2)}}\;.
\end{equation}
Thus
\begin{equation}
   0\le \theta\le\arccos\left[-h\left(\frac{M^2-m_1^2-m_2^2}{2Mm_2\beta\gamma}\right)\right]\;.
\end{equation}

Note that, given any positive values for $M,m_1$ and $m_2$, one can find a sufficiently 
large $\beta$ such that the lower threshold condition Eq.(\ref{TC}) holds. Compare this
with the idea, within the context of the tachyonic neutrino hypothesis, of proton decay in a 
sufficiently boosted frame \cite{CKPG92,ChoKos94}, which has been used
as a way to explain the bend in the knee of the primary cosmic ray spectrum 
\cite{Ehr99a,Ehr99b,Ehr03}. 

\section{Phase space factor for tachyonic case}\label{PhaFac}

The phase space factor relevant for two-body decay for the case of
regular masses $m_1$ and $m_2$ (following the conventions of 
\cite{HugSte90}, who ignore factors of $2\pi$) is 
\begin{eqnarray}
   R_2 &=& \int\int \frac{d^3\,\p_1}{2E_1} \frac{d^3\,\p_2}{2E_2} 
	\delta(M-p_1^0-p_2^0)\delta^{(3)}(\p_1+\p_2) \\
   &=& \frac{\pi p}{M}\;,
\end{eqnarray}
where $p$ is as in Eq.(\ref{eqp1}). 
(Note that in Eq.(11) of \cite{HugSte90} the step function factors $\theta(p_1^0),\theta(p_2^0)$ 
are implied in the product of delta functions $\delta(p_1^2-m_1^2)\delta(p_2^2-m_2^2)$.)
More explanatory details can be found in Ch. 6 of \cite{Gri87}. 

In order to perform the analogous calculation in the case where particle $2$ 
is a tachyon, we write the two-body phase space factor as
\begin{equation}
   R_2 = \int\int d^4\,p_1\delta(p_1^2-m_1^2)\theta(p_1^0)d^4\,p_2\delta(p_2^2+m_2^2)
 	\theta(p_2^0+\beta p_2 x_2)\delta^{(4)}(P-p_1-p_2)\;.
\end{equation} 
Here $P=(M,{\bf 0})$ and the second theta function incorporates the cutoff in the energy-momentum 
spectrum of the tachyon, with the preferred frame assumed to be moving in the $-z$
direction at speed $\beta$ as in Section \ref{ThreCond}. It is expeditious to 
evaluate the quadruple integral with respect 
to the $p_2^\mu$ variables by first integrating over $p_2=\left|\p_2\right|$, the magnitude of the spatial
momentum of $2$. (The quadruple integral with respect to the $p_1^\mu$ variables is integrated first
over $p_1^0$, as usual.) The phase space factor then becomes
\begin{eqnarray}
   R_2 &=& \int\int \left(\frac{d^3\,\p_1}{2E_1}\right)\left(\frac{p_2d\,\phi_2 d\,x_2 d\,p_2^0}{2}
	\theta(p_2^0+\beta p_2 x_2)\right)\delta(M-p_1^0-p_2^0)\delta(\p_1+\p_2) \\
   &=& \frac{\pi p}{2M}\left\lbrace\int_{-1}^1 d\,x_2 \theta(E_2+\beta p_2 x_2)\right\rbrace\label{R2tach}\;,
\end{eqnarray}
where $x_2=\cos\theta_2$ and $\phi_2,\theta_2$ are the azimuthal and polar angles for the spatial 
momentum of $2$, respectively. 

Now one easily sees that if the lower threshold condition Eq.(\ref{TC}) is not satisfied, then the theta
function in Eq.(\ref{R2tach}) is always $0$, giving $R_2=0$. Otherwise, the integral in 
Eq.(\ref{R2tach}) would be a fixed factor between $0$ and $2$ depending only on $M,m_1,m_2$ and $\beta$. 
If the condition 
\begin{equation}
   E_2 \ge \beta p_2\iff E_2\ge\beta\gamma m_2\label{UTC}
\end{equation}
holds, then clearly the theta function is always $1$ on the range of integration, the factor 
in braces in Eq.(\ref{R2tach}) is $2$, and 
\begin{equation}
   R_2=\frac{\pi p}{M}\;. 
\end{equation}
If $-\beta\gamma m_2\le E_2\le \beta\gamma m_2$ then the
theta function in Eq.(\ref{R2tach}) is $1$ when $1\ge x_2\ge -\frac{E_2}{\beta p_2}$, the factor 
in braces is $\{1+\frac{E_2}{\beta p_2}\}$, and 
\begin{equation}
   R_2=\frac{\pi}{2M}\left(p_2+\frac{E_2}{\beta}\right)\;.
\end{equation} 

Note throughout that $p_2=p$ and $E_2$ are evaluated according to Eqs.(\ref{eqp2}) and (\ref{eq4})
respectively. The {\it upper threshold condition} Eq.(\ref{UTC}) is equivalently expressed as 
\begin{equation}
   \left|M-\beta\gamma m_2\right|\ge \sqrt{m_1^2+\gamma^2 m_2^2}\;.
\end{equation} 

\section{$R_2$ as $m_2$ approaches $0$}\label{LimCase}

Our intuition tells us that when $m_2\to 0$ the phase space factors $R_2$ for the regular and tachyonic 
cases must converge to the same thing, namely the phase space factor $R_2$ for a massless particle $2$. 
Let us see how this is borne out in direct calculation. 

The lower threshold condition for the massless case $m_2=0$ is $M\ge m_1$, and a repetition of the 
derivation as for the massive case gives us the formulae
\begin{eqnarray}
   p &=& p_1 = p_2 = \frac{M^2-m_1^2}{2M} \\
   E_1 &=& \frac{M^2+m_1^2}{2M} \\
   E_2 &=& \frac{M^2-m_1^2}{2M}\;,
\end{eqnarray}
which are clearly the limits of Eqs.(\ref{eqp1}), (\ref{eq1}) and (\ref{eq2}) as $m_2\to 0$. 
Furthermore the phase space factor $R_2$ for $m_2=0$ evaluates easily to 
$R_2=\frac{\pi p}{M}=\frac{\pi(M^2-m_1^2)}{2M^2}$, which is the massive case in the limit $m_2\to 0$. 

In order to check that the tachyonic $R_2$ 
approaches the above limit as $m_2\to 0$, we need to verify that in this limit, the factor in braces 
in Eq.(\ref{R2tach}) approaches $2$. If $M=m_1$, then $R_2$ goes to zero as $m_2$ goes to zero (same as the
case of massless $m_2$, where $M=m_1$), thus the factor in braces is irrelevant here. It remains to 
consider the case $M>m_1$. Here, $E_2>0$ for small enough $m_2$, and we can take $m_2$ even smaller, if 
necessary, so that $E_2\ge \beta\gamma m_2$, which, as we have seen, forces the factor in braces to be $2$,
and $R_2$ to be $\frac{\pi p}{M}$, as required. 

As a further observation here, we note that in this last scenario, with $E_2>0$, and $m_2>0$ tachyonic 
and small enough so that $R_2 = \frac{\pi p}{M}$ (i.e., the factor in braces is $2$), the expressions for
regular and tachyonic particle $2$ can be obtained from each other by the replacement 
$m_2^2 \leftrightarrow -m_2^2$. This justifies the assertions made about $R_2$ (as $m_2\to 0$) 
in the introduction.

\section{Conclusions}\label{Conc}

In light of these results, we note some modifications and clarifications that would appear to be 
necessary in the attempt made by \cite{HugSte90} to evaluate the two-body phase space factor. First, with the 
lower cutoff in the energy-momentum spectrum of the tachyon, there would now be a (lower) threshold condition, 
giving an inequality between
$M,m_1,m_2$ and $\beta$ which specifies under what circumstances the decay may proceed at all, Eq.(\ref{TC}). 

Secondly, the phase space factor
$R_2$ (always positive or zero) is, in the tachyonic case, no larger than the expression obtained by replacing 
$m_2^2$ by $-m_2^2$ in the phase space factor for regular $m_2$. When the upper threshold condition Eq.(\ref{UTC}) 
is satisfied, $R_2$ would be {\it exactly} equal to the expression obtained by this replacement. The expression 
Eq.(\ref{R2tach}) thus supercedes the formula Eq.(13) in \cite{HugSte90}. Note that the latter formula must 
also be considered suspect because it gives the wrong limit as $m_2\to 0$.  

Thirdly, note that, according to Eq.(\ref{TC}), if $M^2\ge m_1^2 + m_2^2$, then the lower threshold condition 
would be satisfied 
for any $\beta$ (always positive). Applying this to the case of pion decay, assuming a tachyonic muon (anti-) neutrino
\begin{equation}
   \pi^-\rightarrow \mu^- + \bar\nu_\mu\;,
\end{equation}
this restriction on $m_2=m_{\nu_\mu}$ translates into
\begin{equation}
   m_2 \le \sqrt{M^2-m_1^2} = \sqrt{139.57^2-105.66^2} = 91.19\ {\rm MeV}\;.
\end{equation}
For the simple reason that any violation of this inequality would surely have shown up in pion decay
measurements by now, it seems safe to assume it is satisfied. Then the lower threshold condition is satisfied
for any speed $\beta$ of the preferred frame, and pion decay cannot be prevented in any frame. 

In the paper \cite{HugSte90}, the authors consider the inverse process (when the muon neutrino is tachyonic)
\begin{equation}
   \mu^- \rightarrow \pi^- + \nu_\mu\;.
\end{equation} 
Here, the authors suggest that this decay is always allowed (because they have no lower restriction on the 
energies of the tachyon). In the present approach there {\it is} a restriction: the lower threshold condition 
Eq.(\ref{TC}). Applying it to this case (interchanging $M$ and $m_1$ from the last case), the condition is
\begin{equation}
   \frac{m_2}{2M} + \left(\frac{m_1^2-M^2}{2M}\right)\frac{1}{m_2} \le \beta\gamma\label{RD}\;.
\end{equation}
The minimum value of the LHS is obtained for $m_2=\sqrt{m_1^2-M^2}=91.19$ MeV, and is 
$\frac{\sqrt{m_1^2-M^2}}{M} = 0.863$. Thus $\beta$ must be at least $0.653$ in order to 
allow this decay at all, and that minimum value is achieved for the noticeably large (tachyonic) 
neutrino mass parameter $91.19$ MeV. 
Taking the results of pion decay measurements at face value, the ``worst'' tachyonic value so far obtained
\cite{Ass94} is $m_2 \approx 0.4$ MeV. We plug in this value as an example of how large $\beta$ must be to permit
any ``reverse'' decay to occur: the LHS of Eq.(\ref{RD}) is $98.38$, which implies a $\beta$ of over $0.9999$. 
Thus, it would seem that an unreasonably high $\beta$ would have to exist in order to allow the reverse decay. 

A fourth clarification: in the paragraph before their Eq.(12), the authors of \cite{HugSte90} claim that, 
while for regular particles the restriction
$p^0\ge 0$ must be made to restrict to the positive energy mass shell, the analogous restriction for tachyons
is $p\ge m_2$ (using our notation and conventions). We would postulate that the additional restriction
$p^0+\beta p\cos\theta\ge 0$ must also be made here, and should be considered the analogue of the restriction to the
positive part of the mass shell. The condition $p\ge m_2$, which restricts the magnitude of the three momentum of the 
tachyon, is a very reasonable additional assumption that must be made to rule out imaginary energy modes (which 
lead to observables growing exponentially in time). Indeed in the two-body decay, we have seen that it follows from the
dynamics of the situation under consideration (e.g., conservation of energy and momentum).     
  
In view of the reasonable results we have obtained here for a two-body decay (lower) threshold condition and the
phase space factor $R_2$, we argue that no (egregiously) ``unphysical consequences'' are looming here, and suggest 
that similar sense
can be made of the calculations attempted in other parts of \cite{HugSte90}, according to a QFT incorporating the
cut-off in the energy-momentum spectrum of the tachyon. The prospect of such a suitable theory, which was not considered 
at all in \cite{HugSte90}, would thus at least partly undermine their claim (as stated in their abstract) that 
``there is strong circumstantial evidence against the proposal that at least one neutrino is a tachyon.'' Indeed, 
one may view their ``unphysical consequences'' as arising from a faulty starting assumption about the underlying QFT,
which need not be made (namely that there is no cut-off such as what we have used here). Hence we conclude that 
the results of the present letter at least partly support the tachyonic neutrino hypothesis initially 
made in \cite{CHK85}.  

\noindent
{\bf Acknowledgements:} The author wishes to thank A.S. Wightman for an 
early question about tachyons, and K. Fredenhagen for a similar challenge, 
which provoked me to think about tachyons from a different perspective than usual.
W.G. Unruh and the Dept. of Physics and Astronomy at UBC are acknowledged 
for their extended hospitality, and R. Atkins and B. Kay, for encouragement. 

\bibliography{big}
\end{document}